# Large-Scale Self-Similar Skeletal Structure of the Universe.

V. A. Rantsev-Kartinov

*INF RRC "Kurchatov Institute", Moscow, Russia, rank@nfi.kiae.ru*

## 1. Introduction

The given paper is natural continuation of a series of papers (published earlier [1a, b]) about of a revealing of self-similar skeletal structures (SSSS) in various types of plasma up to space scales. Research by author of the SSSS begin from the analysis of images of various types of plasma by means of a method multilevel dynamic contrasting (MMDC), developed and described earlier [1c, d]. The analysis of images by this method is carried out by imposing of various computer maps of contrasting on the image of plasma received by the various methods and in anyone spectral ranges.

Results of the given analysis of a modern database of images of space objects are shown, that the topology of the revealed space structures is identical to those which have been already found out and described earlier in a wide range of physical environments, the phenomena and scales [1-2]. The typical SSSS consists of separate identical blocks which are linked together to form a network. Two types of such blocks are found: (i) coaxially tubular structures (CTS) with internal radial bonds, and (ii) cartwheel-like structures (CWS), located either on an axle or in the edge of CTS block.

The large-scale skeletal structures of the Universe (SSU) have a whole series of remarkable properties which have been also described before[1c-1f]. So, a long filaments consist of straight ("rigid") nearly identical CTS blocks which is joined flexibly similarly to joints in a skeleton. It is assumed such joints may be realized due to stringing of the individual CTS blocks on common flow of the magnetic field which penetrates the whole such filament, and itself the CTS blocks are an interacting magnetic dipoles with micro-dust skeletons, which are immersed into plasma. Here, the result of the analysis which was been received (with the description of sequence of the lead operations) by means of the MMDC of maps of the Redshift Surveys of galaxies and quasars which have allowed to reveal large-scale structures of the mentioned above topologies are given.

## 2. Observations of similarity of a structuring in a very broad range of length scales

### 2.1. Cartwheel-like structures in the range $10^{-5}$ cm – $10^{23}$ cm.

Here, we will try to draw a bridge between laboratory experiments and space with presenting on consideration of a short gallery of cartwheel-like structures, which are probably the most inconvenient objects for universally describing in the entire range of observed space scales. In a laboratory electric discharges [3,4] and of respective dust deposits [5], the cartwheels are located either in the butt-ends of a tubes or on an «axle-tree» filament, or as a separate block (the smallest cartwheels are of diameter less than 100 nm (see Figs. 2 and 3 in Ref. 5)). So, similar structures of different scales are found in the following typical examples: (i) a big icy particles of a hails (Fig. 1A), (ii) a fragment of tornado (Fig. 1B), (iii) a supernova remnant (Fig. 1C).

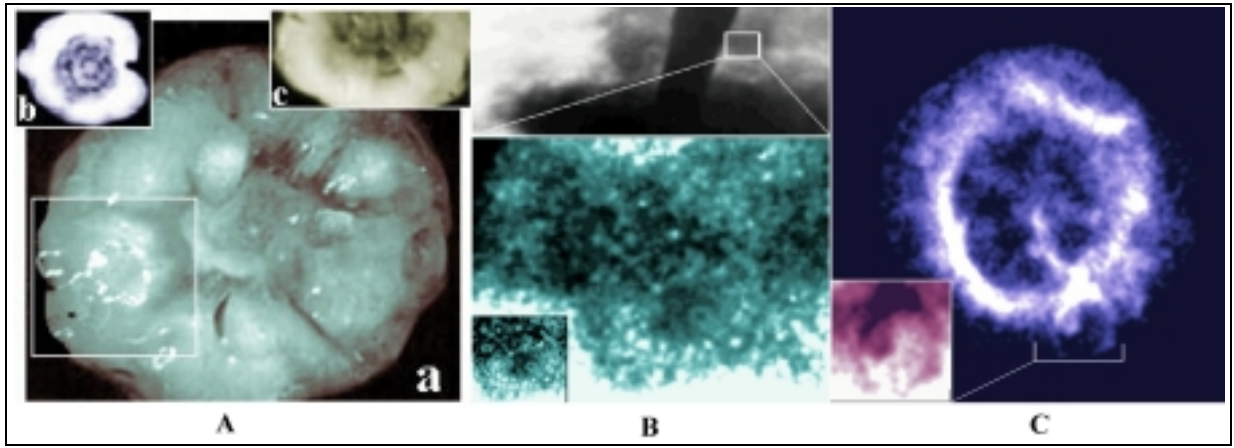

**Fig. 1.** The cartwheel-like structures at different length scales. A, Big icy particles of a hail of diameter 4.5 cm (a), 5 cm (b), and 5 cm (c). The original images are taken from Ref. [6]. The frame in the left lower part of the image (a) is contrasted separately to show an elliptic image of the edge of the radial directed tubular structure. The entire structure seems to contain a number of similar radial blocks. A distinct coaxial structure of the cartwheel type is seen in the central part of image (b). Image (c) shows strong radial bonds between the central point and the «wheel». B, Top section: A fragment of the photographic image [7] of a massive tornado of estimated size of some hundred meters in diameter. Bottom section: A fragment of the top image shows the cartwheel whose slightly elliptic image is seen in the center. The cartwheel seems to be located on a long axle-tree directed down to the right and ended with a bright spot on the axle's edge (see its additionally contrasted image in the left corner insert on the bottom image). C, «A flaming cosmic wheel» of the supernova remnant E0102-72, with «puzzling spoke-like structures in its interior», which is stretched across forty light-years in Small Magellan Cloud, 190,000 light-years from Earth (.../snrg/e0102electricbluet.tif [8]). The radial directed spokes are ended with tubular structures seen on the outer edge of the cartwheel. The inverted (and additionally contrasted) image of the edge of such a tubule (marked with the square bracket) is given in the left corner insert (note that the tubule's edge itself seems to possess a tubular block, of smaller diameter, seen on the bottom of the insert).

Note that the cosmic wheel's skeleton (Fig. 1C) tends to repeat the structure of the icy cartwheel (Fig. 1A) up to details of its constituent blocks. In particular, some of radial directed spokes are ended with a tubular structure seen on the outer edge of the cartwheel. Moreover, in the edge cross-section of this tubular structure, the global cartwheel of the icy particle contains a smaller cartwheel whose axle is directed radial (see left lower window in Fig. 1A). Thus, there is a trend toward self-similarity (the evidences for such a trend in tubular skeletons found in the dust deposits are given in Ref. [5]). Note that the images of Fig. 1 are processed with MMDC [1c,d]. As a rule, the structuring revealed with the help of this method may then be easily recognized in the original, non-processed images (especially, for properly magnified high-resolution images).

Thus, the cosmic wheel's skeleton tends to repeat the structure of the icy cartwheel up to details of its constituent blocks. One may obviously add to the last item of this list namely galaxy - "Cartwheel", which have 150,000 light-years in diameter and 500 million light-years on a distance from Earth in constellation of Sculptor [9].

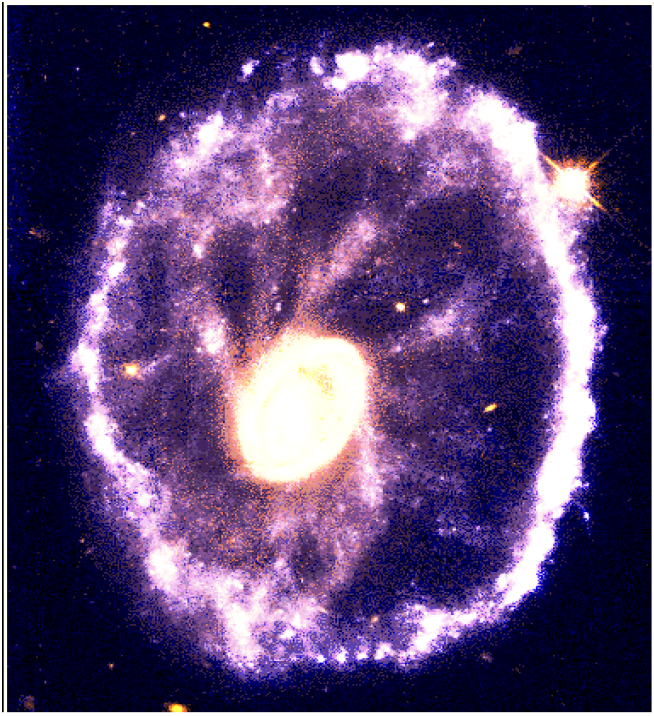

**Fig. 2.** "Cartwheel galaxy" [9], Located 500 million light-years away in the constellation Sculptor, **150,000 light-years across ~ $10^{23}$ cm**, the galaxy looks like a cartwheel. The galaxy's nucleus is the bright object in the center of the image; the spoke-like structures are wisps of material connecting the nucleus to the outer ring of young stars.

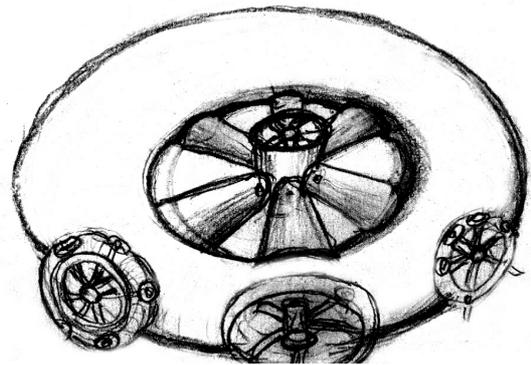

**Fig. 3.** The schematic image of structures such as " cartwheel" is given here. Thus, the cosmic wheel's skeleton tends to repeat the structure of the cartwheel itself up to details of its constituent blocks as in the icy cartwheel.

The wheel-like supernova remnant G11.2-0.3 which have 40 light-years in diameter and 25,000 light-years away in the constellation Sagittarius [10] a two-ring coaxial structure, with the inner ring of one light-year in diameter.

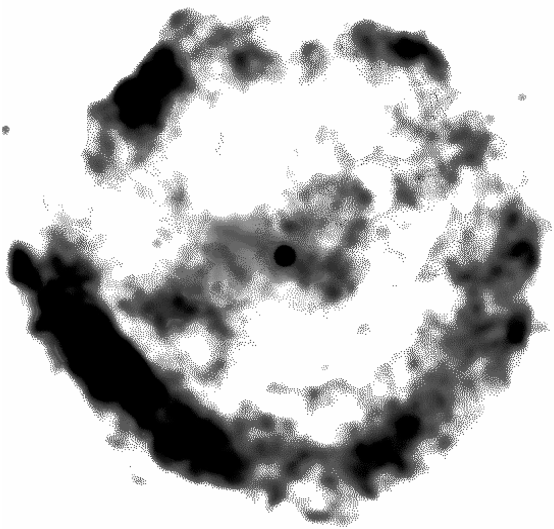

**Fig. 4.** G11.2-0.3**:** A supernova remnant with a central pulsar, located in the constellation of Sagittarius [10], 40 light years across and 25,000 light-years away. Here it is applied MMDC which has allowed to reveal CTS blocks from which the design of given CW (this space object) is made.

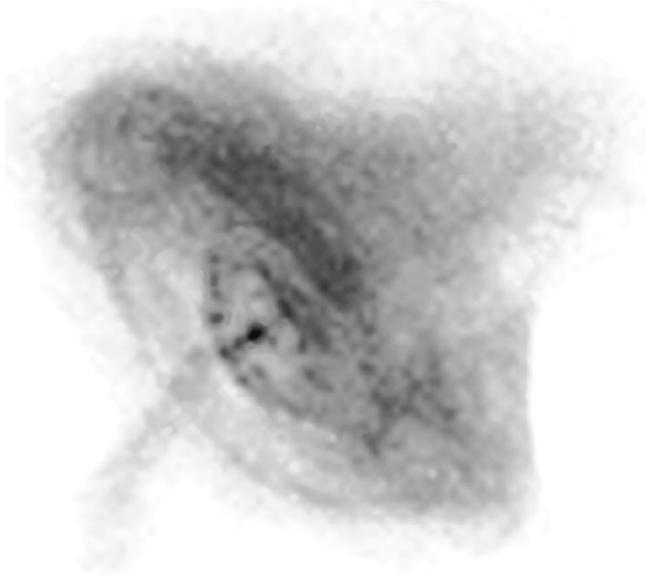

**Fig. 5.** The Crab nebula which is 6,000 light-years away in the constellation Taurus [11]. The Crab Nebula is the remnant of a supernova explosion that was seen on Earth in 1054 AD. It is 6000 light-years from Earth, with the inner ring of one light year ($10^{18}$ cm) in diameter. At the center of the bright nebula is a rapidly spinning neutron star, or pulsar that emits pulses of radiation 30 times a second. Here it is well visible, that the given structure represents system of tubular structures which are telescopic enclosed each into other. At the attentive analysis of the image a radial connections which are characteristic for structure such as CWS are looked through.

## 2.3. The signs of skeletal structuring at cosmological lengths, up to $10^{27}$ cm.

With increasing length scales, the self-illumination of the skeletal network in its certain, critical points continues working but the respective dramatic decrease of the average density of hot radiating baryonic matter leads to observing of exclusively dim dotted imprints of skeletons, like e.g. mysterious dotted images of arcs and circles / ellipses. Note that the typical blocks of skeletons, the cartwheels on an axle and the tubs with the central rod and the cartwheel in the butt-ends are both the dendrites (which are corresponded to examples of the fractal dust deposits - the skeletons composed of tubular nano-fibers [5,12]).

Here it is necessary to note, the blocks of the common network of the Universe (taking into attention of very big distances) can be revealed only in places of infringement of its tightness, i.e. there where takes place breaks up and tears up of network filaments which show a strong luminescence of these areas. Therefore the surface of lengthy filamentary blocks of the Universe either is absolutely not visible, or nevertheless is looked through hardly due to its illumination in places of breakings up of filaments of suitable blocks which surface is formed by these filaments. It appears, that butt-ends of large-scale blocks of the Universe are congestions of galaxies, and galaxies play a role of separate points of these blocks depicting contours. Thus, here the hypothesis is actually offered, that many galaxies can be presented as butt-ends of filamentary tubular blocks which size will be coordinated to scale of galaxies. Cooperating galaxies in that case are butt-ends of the broken blocks or tears of their connections formed in result of cosmology accidents (collisions between filaments and their tears up as a result of tension).

The CWS are the most interesting and complex observable blocks in the Universe, and also they are the most typical blocks of the observable fractal the structures of which are difficultly to confuse with any another. If such structure is well oriented in a flatness of a shearing, then (at condition of a corresponding statistics) the structure clearly becomes apparent because the basic massif of points of this structure is fitted to a rim of a wheel, to its axis and radial spokes, making (on the square) half of area of whole wheel. It allows to identify precisely its under such circumstances. It is theoretically difficult to explain topology CWS by means of magnetic hydrodynamics and the theory of construction of fractals in open systems. The mechanism [1a,b] of construction of the revealed by us topology of fractals spontaneously gathering at formation of electric breakdown at the presence of elementary blocks of a dust, which have tendency to forming structures (for example, as carbon nanotubes or a similar structures but of other elements and chemical compounds) have been earlier considered. The sequence of generations CWS right up to the size ~ $10^{23}$ cm already has been shown earlier [1a,b]. It appears that at largest observable lengths the more or less definite examples of distinctive topology, similar to that of Fig. 1, may be found only in the redshift surveys of *thin* slices of space (the redshift surveys are believed to provide a three dimensional distribution of galaxies, which may give, in particular, the side-on view on a thin conical slice of space). The original data are taken from three different projects, the two-degree-field The result of the analysis means of MMDC [7] of maps of redshifts by together with the image of a kind of structures on an exit of a radial spokes of the CWS at their passage through its rim is given on a next figure.

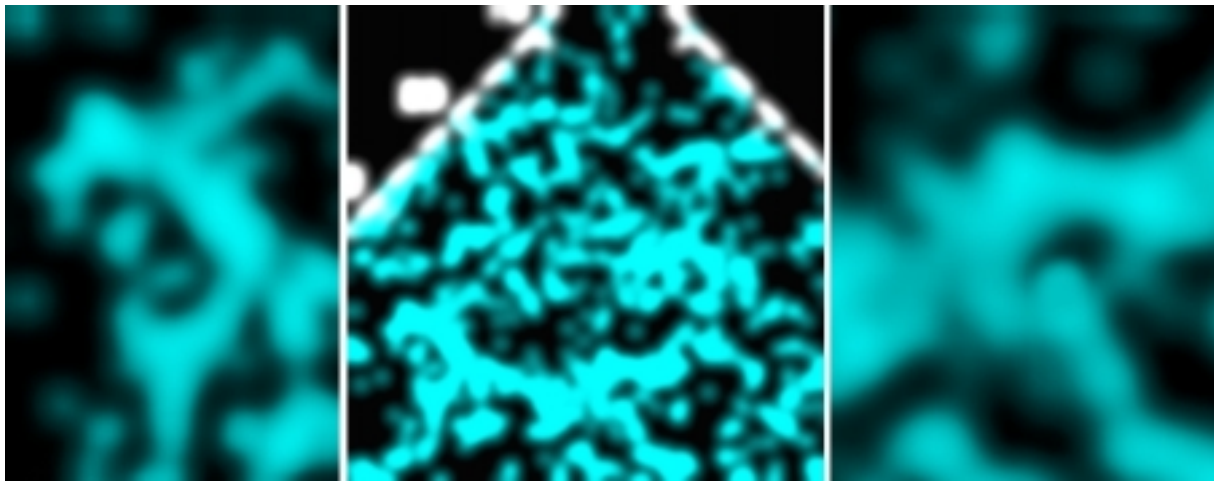

**Fig. 6.** A fragment of similar distribution of the galaxies [13] (20,000 galaxies (for redshifts Z < 0.3, i.e. at distances L up to 2.5 billion light-years away) 1.5° thick slice is cantered at declination -45° in the South Galactic Pole strip, see red points in the colored image at http://www.astro.ucla.edu/~wright/lcrs.html). The left border of the cone crosses the left hand side of the figure at a distance ~ 1.5 $10^9$ light years. Thickening of the spots and subsequent smoothing of the image gives a circle and straight radial filaments.

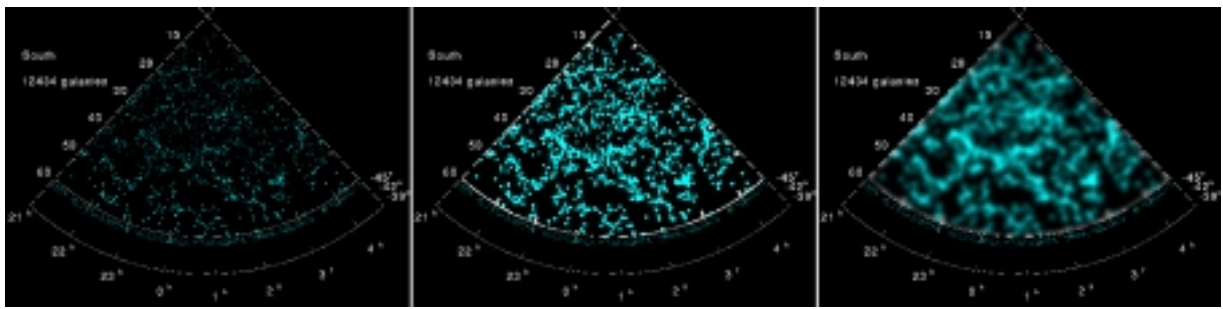

**Fig. 7.** Here the same fragment, as on Fig. 6 is resulted. The left part of figure corresponds to the initial data, average - the same initial data, but all points are increased on the area twice, the right part of figure gives the final image after of Gauss smoothing of intensity distribution of points and carrying out of the correlation analysis by means of MMDC for an establishment of connections between separate points.

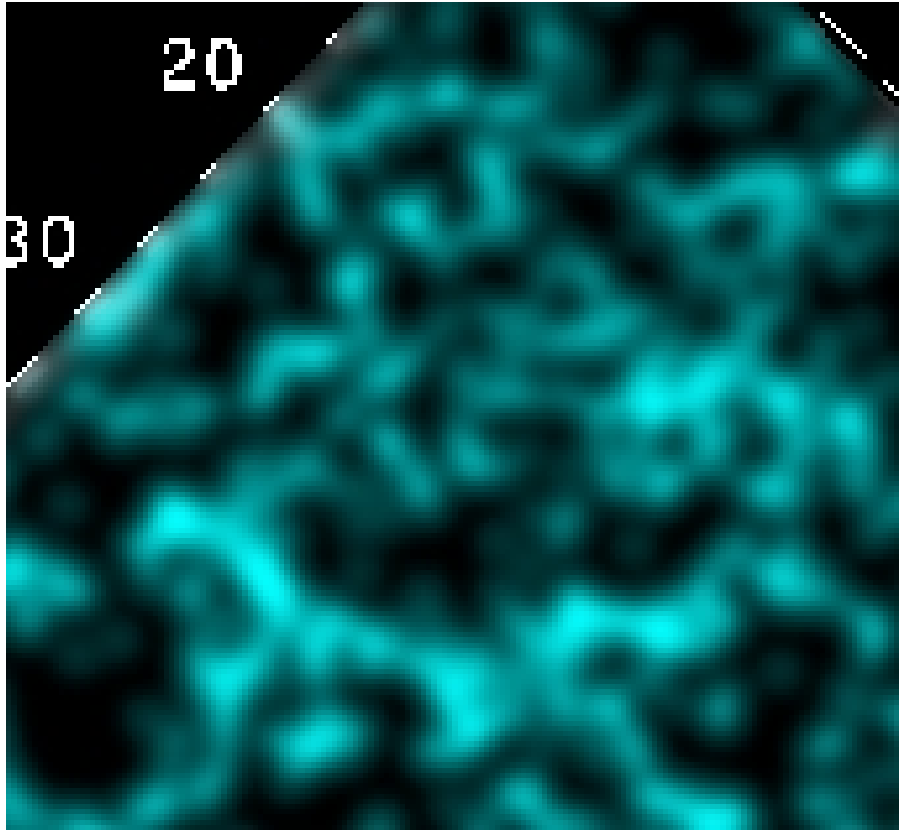

**Fig. 8**. Here the right fragment of Fig. 7 is increased.

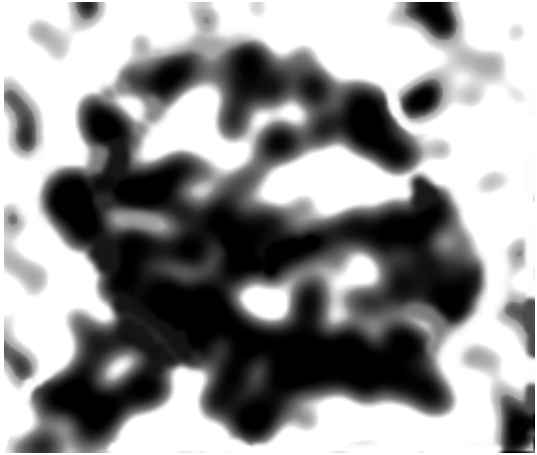

**Fig. 9.** A fragment of similar distribution of the galaxies but some smaller size [13] (1.5° thick slice is centred at declination -6° in the North Galactic Pole strip, see green points in the coloured image at http://www.astro.ucla.edu/~wright/lcrs.html)).

In particular, as the most significant discovery, the author in detail describes the frame structure revealed at it such as CWS in scale ~ $1.5*10^{27}$ cm that makes about 10 % of the scale observable Universe (~1.5 billion light years). The comparative analysis of its structure with similar observable structures of smaller scales is carried out to show their absolute topological identity (Fig. 1).

Let's look now, what structures can be revealed in database of Galaxy Redshift Survey (2dFGRS) [14], which plotted in the redshift space the distribution of some 140,000 galaxies (for redshifts Z <0.3, i.e. at distances L up to 2.5 billion light-years away.

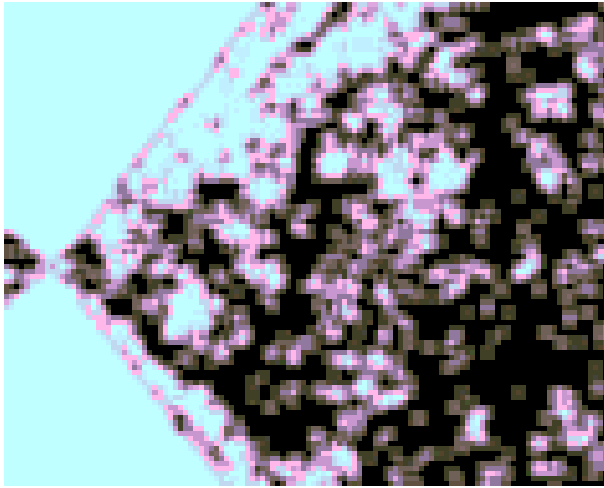

**Fig.10**. A fragment of the projected distribution [14] of the galaxies in the South Galactic Pole strip (4° thick slice is centred at declination -27.5°), as a function of redshift Z and right ascension. The lower border of the cone reaches the bottom of the figure at Z ~ 0.027 (or, equivalently, at a distance ~ 2.7 $10^8$ light years). The slight increase of spots' size in original image at .../Public/Pics/2dFzcone.gif [14] gives elliptic image of a circular (or at least, an arc-like) structure. Despite the structuring seen in this figure is obviously less reliable than that in Fig. 1, the correlation revealed makes it reasonable to suggest an extrapolation of our hypothesis farther to cosmological scales.

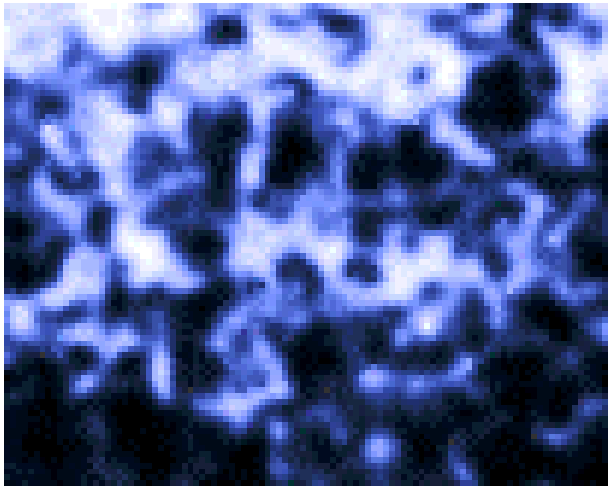

**Fig.11**. An other fragment of the projected distribution [14] of the galaxies. Here the structure such as CWS with its axis located on an axis of this figure is looked through. The rim of this structure is represented as a pentagon. Very easily it is possible to track structure of connection of separate blocks of the common structure and its interweaving into structure of the big scales, i.e., its connections with others, external in relation to it, structures.

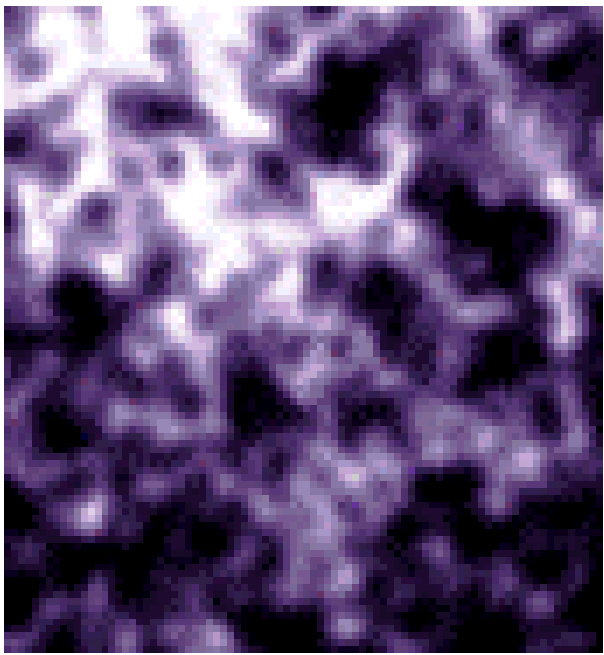

**Fig.12**. One more fragment of the same projected distribution [14] of the galaxies. Here the structure such as CWS with its axis located on an axis of this figure is also looked through. The given structure here is well recognized as well as a perfectly visible connections of it with other blocks of the general structure. The kind and subtleties of connections and structure of separate blocks is well looked through.

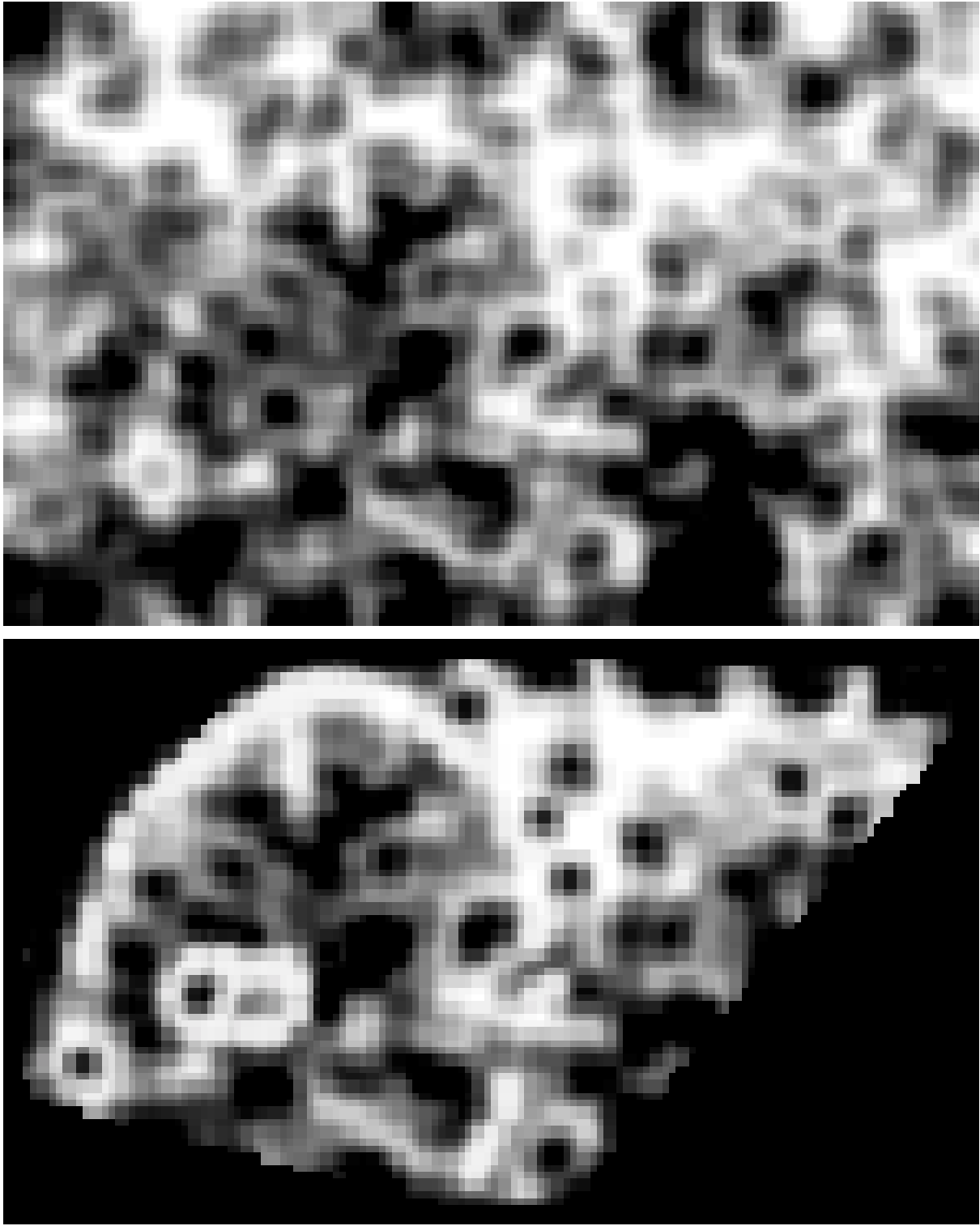

**Fig.13**. One more fragment of the same projected distribution [14] of the galaxies. Here in the lower figure for recognition of its kind it is isolated CTS with structure such as CWS at a forward butt-end of it.

Quasars are the space objects farthest from us. Therefore the structure of their spatial distribution can tell to us about structure of the Universe during very far times. The database of these objects is still small, but, nevertheless, its analysis by means of the described method allows to allocate in an arrangement of quasars the same elements and blocks of a skeletal structures, as for more close galaxies. It can suggest an idea us, that the kind of structure of the universe during those times far from us differed from that structure which we notice and presently a little. Thus, the interpolation of visually correlated sequences of spots of such database (e.g., by means of thickening the spots and subsequent smoothing the image) often gives various skeletal structures (namely, arcs, rings, straight filaments, sometimes the fragments of tubules and cartwheels) of various size and declination with respect to the observer. From optimistic viewpoint, a substantial part of the images may be reduced to a superposition of skeletal structures.

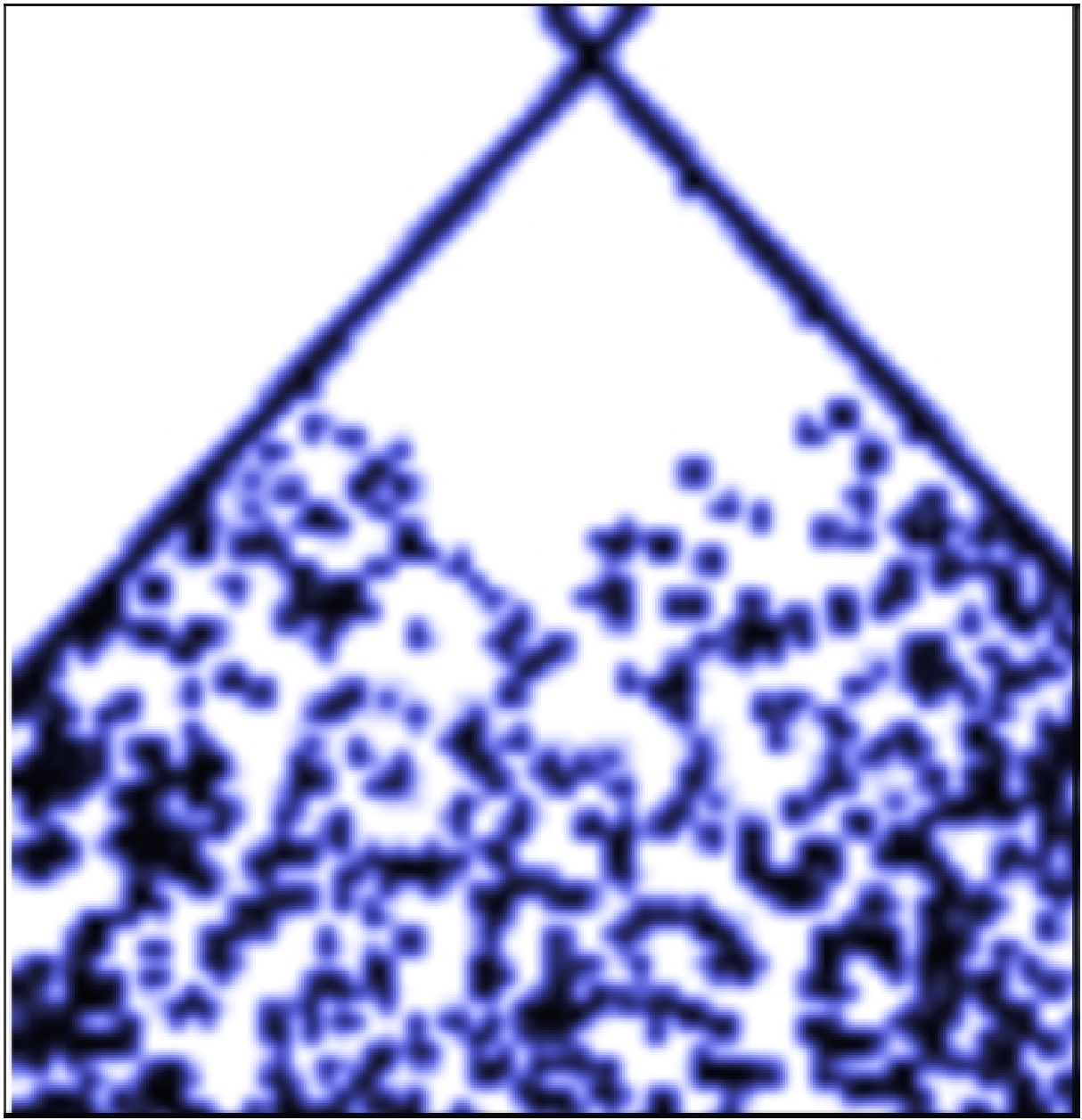

**Fig. 14.** A fragment of the 2dF Quasar Redshift Survey (2QZ) [15], which plotted in the redshift space the distribution of some 20,000 quasars (Z < 3, i.e. L ~ 1.5 $10^{10}$ light-years). Here an each reader can try to unite a separate arches and direct lines, which are designated by points - by quasars in an uniform structure or even into a separate blocks (CWS or CTS) of the corresponding sizes of the Universe.

REFERENCES

[1] A.B.Kukushkin, V.A.Rantsev-Kartinov, a) Phys.Lett. A, **306,** 175, (2002); b) Science in Russia, 1, 42, (2004**);** c) Laser and Particle Beams, **16**, 445,(1998); d) Rev.Sci.Instrum., **70**, 1387,( 1999); e) Current Trends in International Fusion Research: Review and Assessment (Proc. 3[rd] Symposium, Washington D.C., March 1999), Ed. E. Panarella, NRC Research Press, Ottawa, Canada, 121, (2002); f) "Advances in Plasma Phys. Research", (Ed. F. Gerard, Nova Science Publishers, New York), **2**, 1, (2002).
[2] B.N.Kolbasov, A.B.Kukushkin, V.A.Rantsev-Kartinov, et.set., *Phys. Lett. A:*a) **269**, 363, (2000); b) **291,** 447, (2001); c) Plasma Devices and Operations, **8** , 257, (2001).
[3]. Kukushkin, A. B. & Rantsev-Kartinov, V. A. Wild cables in tokamak plasmas. in *Proc. 27-th Eur. Phys. Soc. conf. on Plasma Phys. and Contr. Fusio*n, Budapest, Hungary, June 2000 (http:// epsppd.epfl.ch/Buda/pdf/p2_029.pdf; .../p2_028.pdf).
[4] Kukushkin, A. B. & Rantsev-Kartinov, V. A. Long-lived filaments in fusion plasmas: review of observations and status of hypothesis of microdust-assembled skeletons. In *Current Trends in International Fusion Research, Proc. 4 [th] Symposium,* Washington D.C., 2001 (eds. C.D. Orth, E. Panarella, and R.F. Post) (NRC Research Press, Ottawa, Canada, 2002) (to be published); (see also preprint http://xxx.lanl.gov/pdf/physics/0112091).
[5] Kolbasov, B. N., Kukushkin, A. B., Rantsev-Kartinov, V. A. & Romanov, P. V. Similarity of micro- and macrotubules in tokamak dust and plasma. *Phys. Lett. A,* **26**9, 363-367 (2000).
[6] Australian Severe Weather, http://australiasevereweather.com/photography/ photos/1996/0205mb11.jpg; .../0205mb12.jpg.
[7] National Oceanic & Atmospheric Administration Photo Library, Historic National Weather Service Collection,
http://www.photolib.noaa.gov/historic/nws/images/big/wea00216.jpg.
[8] Chandra X-Ray Observatory - Home. http://chandra.harvard.edu/photo/...
[9] http://hubblesite.org/newscenter/newsdesk/archive/releases/1996/36/image/a
[10] http://chandra.harvard.edu/photo/2001/1227/index.html
[11] http://chandra.harvard.edu/photo/2002/0052/index.html
[12] Kolbasov, B. N., Kukushkin, A. B., Rantsev-Kartinov, V. A. & Romanov, P. V. *Phys. Lett. A,* **29**1, 447-452 (2001).
[13] Shectman, S. A. *et. al.* The Las Campanas redshift survey. *Astrophys. J.* **470**, 172-188 (1996);
[14] The 2dF Galaxy Redshift Survey, http://www.mso.anu.edu.au/2dFGRL/…//
[15] a) The 2dF QSO Redshift Survey, http://www.2dfquasar.org/wedgeplot.html;